\documentclass[twocolumn,aps,prl,showpacs,,showkeys,groupedaddress,amsmath,amssymb]{revtex4}
\usepackage{amsmath}
\usepackage{amssymb}
\usepackage{txfonts}
\usepackage{graphicx}
\usepackage{dcolumn}

\begin{document}

\oddsidemargin=-10mm

\title{Robust quantum repeater with atomic ensembles and single-photon sources}
\author{Fang-Yu Hong}
\author{Shi-Jie Xiong }

\affiliation{National Laboratory of Solid State Microstructures and
Department of Physics, Nanjing University, Nanjing 210093, China}
\date{\today}
\begin{abstract}
 We present a quantum repeater protocol using atomic ensembles, linear optics and single-photon sources. Two  local 'polarization' entangled states of atomic ensembles $u$ and $d$ are generated by absorbing a single photon emitted by an on-demand single-photon sources, based on which high-fidelity local entanglement between four ensembles can be established efficiently through Bell-state measurement. Entanglement in basic links  and entanglement connection between links are carried out by the use of two-photon interference. In addition to  being robust against phase fluctuations in the quantum channels, this scheme may speed up quantum communication with higher fidelity by about 2 orders of magnitude for 1280 km compared with the partial read (PR) protocol  (Sangouard {\it et al.}, Phys. Rev. A {\bf77}, 062301 (2008)) which may generate entanglement most quickly among the previous schemes with the same ingredients.
\end{abstract}


\maketitle

Entanglement plays a fundamental role in quantum information science
\cite{pzol} because it is a crucial requisite for quantum metrology
\cite{vgsl}, quantum computation \cite{jcpz,ldhk}, and quantum
communication \cite{jcpz,hbwd}. Because of losses and other noises in
quantum channels, the communication fidelity falls exponentially
with the channel length. In principle, this problem can be
overcome by applying quantum repeaters \cite{hbwd,lcjt, llsy,
llnm, wmrm, fyhsx}, in which initial imperfect entangled pairs are established over elementary links, these initial pairs are then purified to high fidelity entanglement and connected through quantum swaps \cite{chbe,zzhe} with doubled quantum communication length. With the quantum repeater protocol one may generate high fidelity long-distance entanglement with resources increasing only polynomially with communication distance. A protocol of special importance for
long-distance quantum communication with atomic ensemble as local memory qubits and linear optics  is proposed in a seminal paper of Duan {\it
et al.} \cite {dlcz}. After that considerable efforts have been
devoted along this line \cite{crfp,cldc,kchd,ssmz,sras,zccs, zyyc}.

In additional to using relatively simple ingredients, DLCZ protocol has built-in entanglement purification and thus is tolerant against photon losses.  However, entanglement generation and entanglement connection in the DLCZ protocol is based on a single-photon Mach-Zehnder-type interference, resulting the relative phase in the entangled state of two distant ensembles is very sensitive to path length instabilities \cite{zccs, zcbz}. Moreover, entanglement generation is created by
detecting a single photon from one of two ensembles. The probability
 of generating one excitation in two ensembles denoted by $p$ is related to the
fidelity of the entanglement, leading to the condition $p\ll 1$
to guaranty an acceptable quality of the entanglement. But when
$p\rightarrow0$, some experimental  imperfections such as stray
light scattering and detector dark counts will contaminate the
entangled state increasingly \cite{zyyc}, and subsequent processes
including quantum swap and quantum communication become more
challenging for finite coherent time of quantum memory \cite{kchd}.

In  recent papers \cite{zccs, zcbz},  Chen {\it et al.} suggest a robust quantum repeater protocol which is insensitive to the path length phase instabilities by using  the two-photon Hong-Ou-Mandel-type (HOMT) interference rather than single-photon interference. Sangouard {\it et al.} \cite{nscsb} developed that protocol by exploiting a more efficient method of generating entangled pairs locally with partial readout  of the ensemble-based memories, which makes it generate entanglement most quickly among the previous schemes with the same ingredient.  However, to achieve a high communication rate atomic ensembles have to be excited with a very high repetition rate because of the very low probability
$p$. Here we propose a quantum repeater strategy based on the DLCZ scheme. In this strategy, the local entangled state between two atomic ensembles is established by storage a sharing photon from an on-demand  single-photon source \cite{kchd, nscs, fhsx}, which may release the stringent condition on $p$, both of initial entanglement states in basic links  and entanglement connection between links are carried out through detection of the two-photon HOMT interference. Besides being insensitive to phase fluctuations in the quantum channels, this scheme may enhance quantum communication with near unity fidelity by about 2 orders of magnitude for 1280 km compared with the PR protocol.

\begin{figure}
\includegraphics[width=0.8\columnwidth]{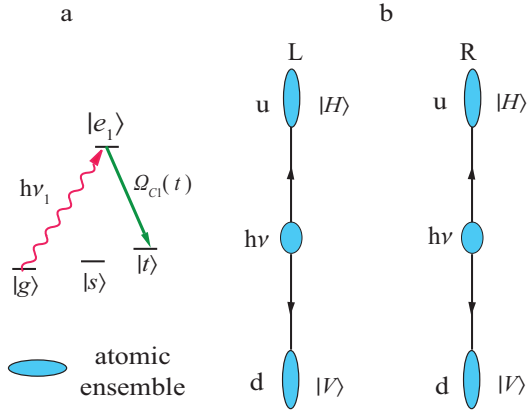}
\caption{\label{fig:1}(Color online) (a) The relevant level
configuration of atoms in the ensembles and the coupling to
pulses. (b) Schematic illustration of
entanglement establishment between two atomic ensembles $u$ and $d$ via  ensembles coherent absorbing a shared photon from an on-demand single-photon sources.  }
\end{figure}

An atomic ensemble consists of a cloud of $N_a$ identical
atoms with pertinent level structure shown in Fig. \ref{fig:1} a.
One ground state $|g\rangle$ and two  metastable states $|s\rangle$
and $|t\rangle$ may be provided by, for instance, hyperfine or
Zeeman sublevels of the electronic ground state of alkali-metal
atoms, where long relevant coherent lifetime has been observed
\cite{jhjs, dpaf, twsc}. The atomic ensemble is optically thick
along one direction to enhance the coupling to light \cite{dlcz}.
State $|e_1\rangle$ is an excited state.  Let us consider two sites $L$ and $R$ at every node shown in Fig.\ref{fig:1}. Each site has two atomic ensembles $u$ and $d$ acting as one memory qubit.  A single photon emitted with a repetition rate $r$
from an on-demand single-photon source \cite{kchd, sfas} located
halfway between atomic ensembles $u$  and $d$ is split into an entangled
state of optical modes $u_{in}$ and $d_{in}$ (Fig. \ref{fig:1} b)
\begin{equation}\label{eq4}
  |\psi_{in}(\phi)\rangle= \frac{1}{\sqrt{2}} \left(|0 \rangle_{u_{in}}|
  1\rangle_{d_{in}}+{\text e}^{i\phi}|1\rangle_{u_{in}}|0\rangle_{d_{in}}\right)
\end{equation}
 where $\phi$ denotes an unknown difference of the phase shifts in the $u$ and $d$ side channels.
This state then is coherently  mapped onto the state of atomic
ensembles $u$ and $d$:
\begin{equation}\label{eq5}
|\psi(\phi)\rangle_{ud}=  \frac{1}{\sqrt{2}}\left(T_u^\dagger+{\text e}^{i\phi}T_d^\dagger\right)|0_a\rangle_u|0_a\rangle_d
 \end{equation}
by applying techniques such as adiabatic passage based on dynamic
electromagnetically induced transparency \cite{kchd}, where
$T\equiv1/\sqrt{N_a}\sum_{i=1}^{N_a}|g\rangle_i\langle t|$ is the
annihilation  operator for the symmetric collective atomic mode $T$
\cite{dlcz} and $|0_a\rangle\equiv\otimes_i|g\rangle_i$ is the
ensemble ground state. Considering  the inefficiency of the excitation transfer from the optical mode to quantum memory
mode, the state of memory qubits can be described by an
effective maximally entangled (EME) state \cite{dlcz}
 \begin{equation}\label{eq7}
    \rho_{ud}(\phi)=(1-\eta_p\eta_s)|0_a0_a\rangle_{ud}\langle0_a0_a|+
    \eta_p\eta_s|\psi(\phi)\rangle_{ud}\langle\psi(\phi)|
 \end{equation}
 where $\eta_p$ denotes the probability of emitting one photon by the
single-photon source per pulse, $\eta_s$ is the efficiency of successful storing a photon in an atomic ensemble and  $(1-\eta_p\eta_s)$ is the vacuum coefficient.

\begin{figure}
\includegraphics[width=0.8\columnwidth]{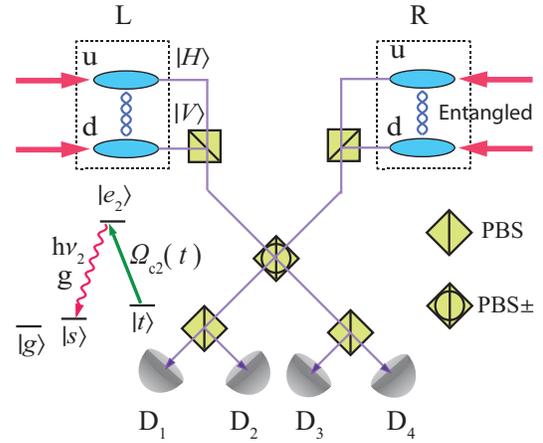}
\caption{\label{fig:2}(Color online) Schematic illustration of local
entanglement generation at every nodes of a quantum channel. The memory qubits at sites $L$, $R$ are prepared in the  entangled states in the form of equation \eqref{eq7}. With the memory qubits at sites $R$ and $L$ illuminated by  retrieve pulses near resonant with the transition $|t\rangle\rightarrow|e_2\rangle$ the anti-Stokes photons are generated with different polarizations $|H\rangle$ from $u$ ensembles and $|V\rangle$ from $d$ ensembles,  and subject to BSM at the midpoint. Up to a local unitary phase shift the coincidence count between single-photon detectors $D_1$ and $D_4$ ($D_1$ and $D_3$) or $D_2$ and $D_3$ ($D_2$ and $D_4$ will project the memory qubits into a PME state between sites $L$ and $R$ in the form of equation \eqref{eq1}. PBS (PBS$_\pm$) transmits  $|H\rangle$ ($|+\rangle$) photons and reflects  $|V\rangle$ ($|-\rangle$) photons, where $|\pm\rangle=\frac{1}{\sqrt{2}}(|H\rangle \pm |V\rangle)$.}
\end{figure}

Before proceeding we discuss the conversion of the collective atomic excitation  $T$ into the atomic excitation $S$ given by
$S\equiv1/\sqrt{N_a}\sum_{i=1}^{N_a}|g\rangle_i\langle s|$.
 Consider the atoms have an excited
state $|e_2\rangle$ fulfilling the condition that the dipole moments of the atomic transitions $
e\textbf{r}_{1}=e\langle g|\textbf{r}|e_2\rangle=0$, $
e\textbf{r}_{2}=e\langle s|\textbf{r}|e_2\rangle\neq0$, and $
e\textbf{r}_{3}=e\langle t|\textbf{r}|e_2\rangle\neq0$ \cite{mfai}. The transition $|s\rangle\rightarrow|e_2\rangle$ of each of these atoms is coupled to a quantized radiation mode  with a coupling constant $g$; the transitions from $|e_2\rangle\rightarrow|t\rangle$ are resonantly driven by a classical control field of Rabi frequency $\Omega_{c2}$ (Fig.\ref{fig:2}). The dynamics of this systems can be described by the interaction Hamiltonian \cite{mlsy}
\begin{equation}\label{eq8}
 H_{in}=\hbar g \hat{a}\sum_{i=1}^{N} \hat{\sigma}_{e_2s}^i+\hbar \Omega_{c2}(t)\sum_{i=1}^{N} \hat{\sigma}_{e_2t}^i +H.c.
\end{equation}
where $\hat{a}$ is the annihilation operator of the quantized radiation mode and $\sigma_{\mu\nu}^i=|\mu\rangle_{ii}\langle\nu|$ is the flip operator of the $i$th atom between states $|\mu\rangle$ and $|\nu\rangle$. This interaction Hamiltonian has the dark state with zero adiabatic eigenvalue \cite{appm, mlsy, mfml},
\begin{equation}\label{eq8}
    |D\rangle=\cos\theta(t) S^\dag|g\rangle|1\rangle-\sin \theta(t)T^\dag |g\rangle|0\rangle
\end{equation}
where $\tan\theta=g/\Omega_{c2}(t)$ and $|n\rangle$ denotes the radiation state with $n$ photon. Thus based on this dark state,
by applying a retrieval pulse of suitable polarization that is
resonant with the atomic transition
$|t\rangle\rightarrow|e_2\rangle$, the atomic excitation $T$ in an
atom ensemble can be converted into the atomic excitation $S$ while an anti-Stokes
photon which has polarization and frequency different from the
retrieval pulse is emitted  \cite{dlcz,mlsy,dpaf,mfai,clzd}.

The two memory qubits at $L$ and $R$ are prepared in the state $
\rho_{ud}(\phi)$, then illuminated simultaneously by retrieval  laser pulses on resonance of the atomic transition
$|t\rangle\rightarrow|e_2\rangle$, the atomic excitations $T$ are
transformed simultaneously into excitations $S$ while anti-Stokes photons are
emitted. We assume the anti-stokes photons produced from the memory qubits are in an orthogonal polarization state $|H\rangle$ from ensemble $u$ and $|V\rangle$ from ensemble $d$, which represent horizontal and vertical linear polarization, respectively, resulting in an entangled state of the memory qubit and the anti-Stokes photon.

  After the conversion, the stokes photons from site $L$ and $R$ at every node are directed to the polarization beam splitter (PBS) and experience two-photon Bell-state measurement (BSM) (shown in Fig.\ref{fig:2}) at the middle point to generate an entanglement between the two memory qubits $L$ and $R$.
   Only the
coincidences of the two single-photon detectors $D_1$ and  $D_4$ ($D_1$ and  $D_3$) or $D_2$ and  $D_3$ ($D_2$ and  $D_4$) are recorded, so the protocol
is successful only if each of the paired detectors have a click. Under this circumstance, the
vacuum components in the EME states, one-excitation components like $S^\dag_{L_u}|vac\rangle$, and the two-excitation components
$T^\dag_{L_u}T^\dag_{R_d}|vac\rangle$ and
$T^\dag_{L_d}T^\dag_{R_u}|vac\rangle$ have no effect on the
experimental results, where $|vac\rangle$ is the ground state of the
ensemble $|0_a\rangle_{u_L}|0_a\rangle_{d_L}|0_a\rangle_{u_R}|0_a\rangle_{d_R}$. A coincidence click between single-photon detectors,for example, $D_1$ and  $D_4$  will project the two memory qubits into the polarization maximally entangled (PME) state \cite{zcbz,qzxb}
  \begin{equation}\label{eq1}
    |\Psi\rangle_{LR}=\frac{1}{\sqrt{2}}(S^\dag_{u_L}S^\dag_{u_R}+S^\dag_{d_L}S^\dag_{d_R})|vac\rangle.
  \end{equation}
    The success probability for
entanglement generation at every node is $p_l=\eta_p^2\eta_s^2\eta_{e_1}^2\eta_d^2/2$, where $\eta_{e_1}$ denotes the efficiency for the
atomic ensemble emitting a photon during the process $T^\dag|0_a\rangle\rightarrow S^\dag|0_a\rangle$ and $\eta_d$ denotes  the single-photon detection efficiency. The average waiting time for successful generating a local entanglement state is $T_l=\frac{1}{rp_l}$.

\begin{figure}
\includegraphics[width=0.8\columnwidth]{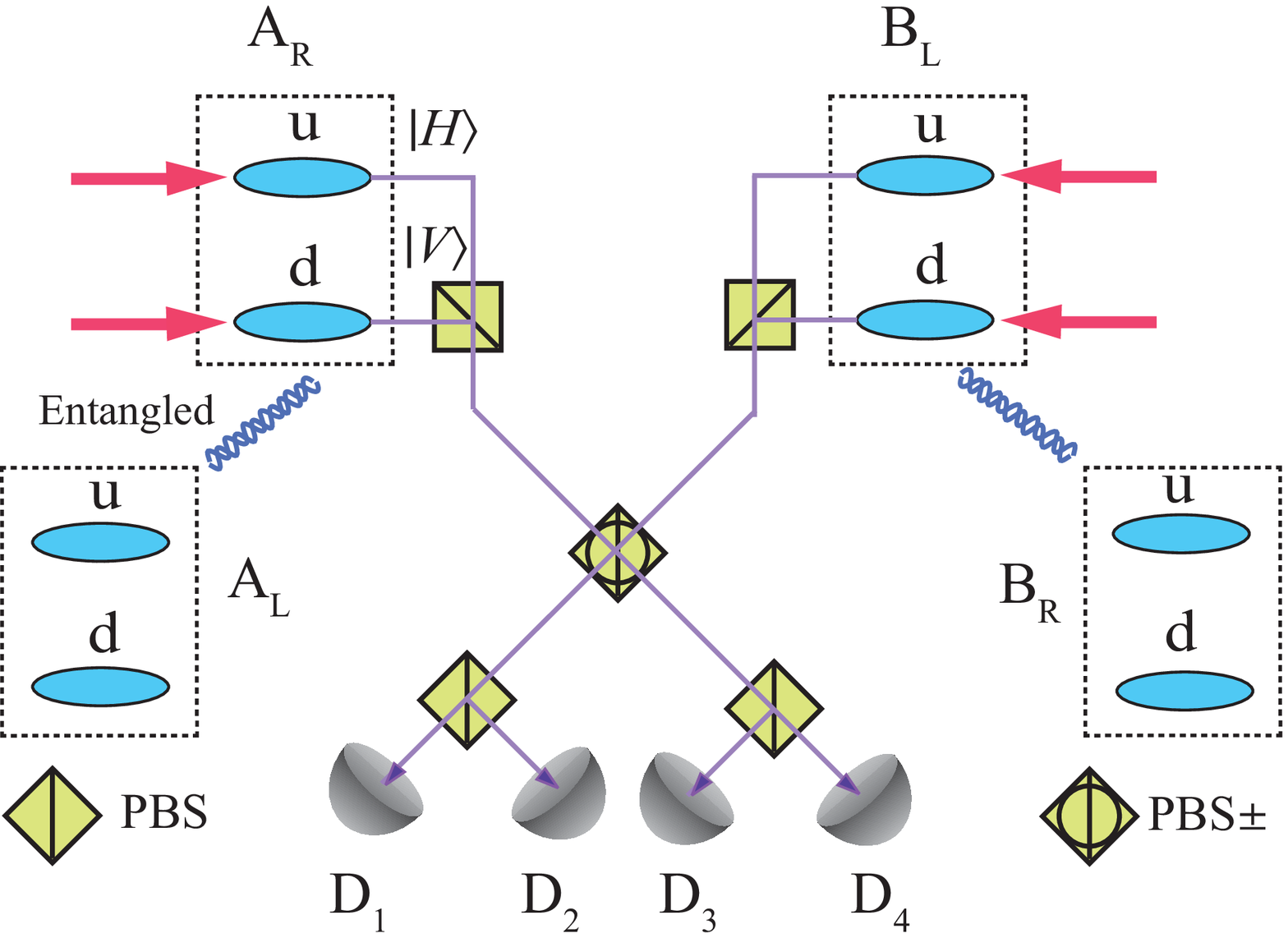}
\caption{\label{fig:3}(Color online) Schematic illustration of
entanglement generation  between  two nodes $A$ and $B$ via BMS. The memory qubits at sites ($A_L$, $A_R$) and  ($B_L$,  $B_R$) are prepared in the PME states in the form of equation \eqref{eq1}. With the memory qubits at sites $A_R$ and $B_L$ illuminated by near resonant retrieve pulses the anti-Stokes photons are generated with different polarizations $|H\rangle$ from $u$ ensembles and $|V\rangle$ from $d$ ensembles,  and subject to BSM at the midpoint. Up to a local unitary phase shift the coincident count between $D_1$ and $D_4$ ($D_1$ and $D_3$) or $D_2$ and $D_3$ ($D_2$ and $D_4$) will project the memory qubits into a PME state between sites $A_L$ and $B_R$.}
\end{figure}

After local entanglement states are established, one can generated entangled state in a basic link with nodes $A$ and $B$ at a distance $L_0$ using BSM illustrated in Fig. \ref{fig:3}. By
applying a retrieval pulse of suitable polarization that is
near-resonant to the atomic transition
$|s\rangle\rightarrow|e_3\rangle$, the atomic excitation $S$ in the
atom ensemble can be converted into light which has polarization and
frequency different from the retrieval pulse.  The efficiency of this transfer denoted by $\eta_{e2}$ can be very close to unity because of collective enhancement \cite{dlcz, mlsy,dpaf,mfai,clzd} When four atomic ensembles are illuminated by the retrieval pulses, the anti-stokes photons from memory qubits at $A_R$ and $B_L$ are directed to the polarization beam splitter (PBS) and subject to BSM. The coincident clicks between single-photon detectors $D_1$ and  $D_4$ ($D_1$ and  $D_3$) or $D_2$ and  $D_3$ ($D_2$ and  $D_4$) will project the two memory qubits into the PME state
  \begin{equation}\label{eq2}
    |\Psi\rangle_{AB}=\frac{1}{\sqrt{2}}(S^\dag_{A_{L_u}}S^\dag_{B_{R_u}}+S^\dag_{A_{L_d}}S^\dag_{B_{R_d}})|vac\rangle
  \end{equation}
  up to a local unitary transformation \cite{zcbz,qzxb}. The protocol is successful if, and only if these coincident clicks occur with a  probability $p_0=\eta_{e_2}^2\eta_d^2\eta_t^2/2$, where  $\eta_t=\text{exp}[-L_0/(2L_{att})]$ is the fiber transmission  with the attenuation length $L_{att}$ .

\begin{figure}
\includegraphics[width=0.8\columnwidth]{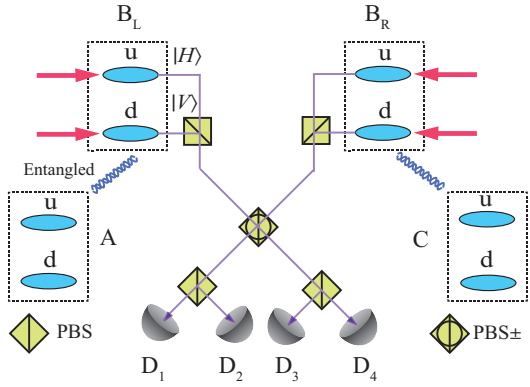}
\caption{\label{fig:4}(Color online) Schematic illustration of
entanglement connection  between  two nodes $A$ and $C$ via entanglement swapping. The memory qubits at sites ($A$, $B_L$) and  ($B_R$,  $C$) are prepared in the PME states in the form of equation \eqref{eq2}. With the memory qubits at sites $B_L$ and $B_R$ illuminated by near resonant retrieve pulses the anti-Stokes photons are generated with different polarizations $|H\rangle$ from $u$ ensembles and $|V\rangle$ from $d$ ensembles,  and subject to BSM at the midpoint. Up to a local unitary phase shift the coincidence count between $D_1$ and $D_4$ ($D_1$ and $D_3$) or $D_2$ and $D_3$ ($D_2$ and $D_4$ will project the memory qubits at sites $A$ and $C$ into a PME state with a double communication length.}
\end{figure}

After successful generating entanglement within basic links, we
can extend the quantum communication distance through entanglement
swapping with the configuration shown in Fig.\ref{fig:4}.  We
have two pairs of ensembles---$A_{u}$, $A_{d}$,
$B_{L_u}$, and $B_{L_d}$, and
$B_{R_u}$, $B_{R_d}$,
$C_{u}$, and $C_{d}$, located at three sites $A$, $B$, and $C$. Each pair of ensembles is prepared in the PME state in the form of Eq.\eqref{eq2}. The stored
atomic excitations of four neighboring atomic ensembles  $B_{L_u}$, and $B_{L_d}$, and
$B_{R_u}$, $B_{R_d}$ are transferred into light simultaneously with the retrieve pulses. We also assume the polarizations of the anti-stokes photons produced from the $u$ ensemble and $d$ ensemble are orthogonal. The stimulated optical excitations are then subject to BSM at the middle point. If, and only if coincident clicks between
detectors $D_1$ and  $D_4$ ($D_1$ and  $D_3$) or $D_2$ and  $D_3$ ($D_2$ and  $D_4$) occur, the
protocol is successful with a probability $p_1=\frac{1}{2}\eta_{e_2}^2\eta_d^2$ and an entangled state in
the form of equation \eqref{eq2} is established among the ensembles
$A_u$, $A_d$, $C_u$, and $C_d$ with a doubled communication
distance. Otherwise, we need to repeat the previous process of
entanglement generation and swapping.

The scheme for entanglement swapping can be applied to arbitrarily
extend the communication distance. For the $i$th ($i=1,2,...,n$)
entanglement swapping, we first prepare simultaneously two pairs of
ensembles in the PME state  with the same communication length
$L_{i-1}$, and then make entanglement swapping as shown by Fig.\ref{fig:4} with a success probability $p_i=\frac{1}{2}\eta_{e_2}^2\eta_d^2$. After a successful
entanglement swapping, a new PME state is established and  the
communication length is extended to $L_i=2L_{i-1}$. Since the $i$th
entanglement swapping needs to be repeated on average $1/p_i$ times,
the average total time needed to generating a PME state over the
distance $L=L_n=2^nL_0$ is given by
 \begin{equation}\label{eq6}
T_{tot}=\frac{T_0}{\prod_{i=0}^np_i}
\end{equation}
where $T_0= L_0/c +T_l$ with $c$ being the light speed in the optical
fiber.

 For the prototype two-photon-based protocol and the PR protocol, the established local entangled states are mixed states due to higher-order excitations in the atomic ensembles. Thanks to applying on-demand single-photon sources, the higher-order excitations can be arbitrarily suppressed with unending advances in single-photon
sources \cite{sfas,blmo}, resulting the fidelity of local entanglement and the final long-distance entanglement approaching unity when there are no other imperfections.
 In our scheme, the probability of generating an atomic excitation via absorbing a photon from an on-demand single-photon sources does not have to meet the condition $p\ll 1$. Considering the efficiency of the transfer of $T$ into $S$ $\eta_{e_1}$ may be low, we assume $\eta_{e_1}=0.05$.    Assuming that $r=39.2$ MHz, $\eta_p=0.9$, $\eta_{s}=0.9$, $\eta_{e_2}=0.9$,
$\eta_d=0.9$, $L=1280$ km, $L_{att}=22$ km for photons with
wavelength of $1.5\, \mu$m \cite{nscs}, $c=2.0\times10^5$ km/s, and
$n=4$, equation (\ref{eq6}) gives the average total time
$T_{tot}=4.4$ s. For comparison reason , we estimate  the average total time
 $T_{tot}=107.6$ s for the
PR protocol \cite{nscsb} with  the above relevant
parameters in addition to the probability $p=0.006$ to obtain entanglement fidelity $F=0.9$ \cite{nscsb}. With the above parameters, to equate local entanglement preparation time $T_l$ with the communication time $L_0/c$, we have the repetition rate $r=3.76$ MHz for this new scheme and $r=39.2$ MHz for the PR protocol.  Note that the possibility of  high repetition rate of pulse acting on atomic ensembles remains an open question, since only a repetition rate $r=250$ KHz is reported by Chou {\it et al.} \cite{ccsp}. As for this new scheme, the requirement of repetition rate $r$ is weak, at the same time we note that $\eta_{e_1}$ can be enhanced by putting the atomic ensembles in a low-finesse ring cavity \cite{dlcz} and  one can exploited many kinds of on-demand single-photon sources, such as molecule-based sources with max rate $100$ MHz and quantum-dot-based sources with max rate 1 GHz \cite{blmo}. In this scheme, average total time $T_{tot}=0.84$ s can be reached with optimal number of links $n=6$ and the same  aforementioned relevant parameters for $L=1280$ km. Thus through this scheme quantum communication with higher fidelity may be sped up by about two orders of magnitude   for 1280 km compared with that based on the PR protocol.

Now we discuss imperfections in our architecture for quantum
entanglement distribution.  We have shown that this strategy has no inherent error mechanism, that is, the fidelity of the obtained entanglement will be unity provided that all components of the setup work perfectly. In the whole process of entanglement
generation, connection, the photon
losses includes contributions from channel absorption, spontaneous
emissions in atomic ensembles, conversion inefficiency of
single-photon into and out of atomic ensembles, and inefficiency of
single-photon detectors. These losses decreases the success probability
but has no effect on the fidelity of the established entanglement. Main imperfection is due to dark counts, which means that detector clicks in the absence of photons. When a dark count occurs in  either the stage of basic entanglement generation or that of entanglement connection, one can exploits photon-number-resolved single-photon detector to exclude this case according to repeater protocol. If a dark count occurs in the process of local entanglement preparation, the local entangled state is a mixed state with a contribution from one atomic excitation, resulting the fidelity of the generated long-distance entanglement decreases. Considering that the probability  for a detector to give a dark count denoted by $p_d$ smaller than $5\times10^{-6}$ is within the reach of the current techniques \cite{nscs}, we can estimate the fidelity imperfection $\Delta F\equiv1-F$ for the generated long-distance entanglement by
\begin{equation}\label{eq3}
    \Delta F=2^{n+1}p_d<1.6\times 10^{-4}
\end{equation}
  for $n=4$. Further note that this scheme is compatible with the linear optics entanglement purification protocol introduced in the manuscript \cite{jpcs}.

In conclusion, we  have proposed a scheme for long-distance entanglement distribution based on two-photon-interference and single-photon sources.
Through this scheme, the rate of long-distance quantum communication
can increase by about two orders of magnitude for 1280 km compared
with the PR protocol. At the same time, this scheme is robust against path length instabilities and near
unity fidelity of generated entanglement may be
expected. Considering the simplicity of the physical set-ups used,
this scheme may opens up the probability of efficient long-distance
quantum communication.

{\it Acknowledgments} This work was supported by the State Key
Programs for Basic Research of China (2005CB623605 and
2006CB921803), and by National Foundation of Natural Science in
China Grant Nos. 10874071 and 60676056.

\end{document}